\documentclass[aip,jap,reprint]{revtex4-1}
\usepackage{amsmath}
\usepackage{amsfonts}
\usepackage{amssymb}
\usepackage{graphics}
\usepackage{float}
\usepackage{graphicx}
\usepackage{epstopdf}
\usepackage[compact]{titlesec}
\usepackage{natbib}
\usepackage{threeparttable}
\usepackage[titletoc,title]{appendix}
\usepackage{lipsum}
\usepackage{xcolor}
\usepackage{subfigure}
\usepackage{color}

\newcommand{\etal}{\textit{et al}. }
\newcommand{\ie}{\textit{i}.\textit{e}., }

\begin{document}
\title{Low-temperature thermal transport and thermopower of monolayer transition metal dichalcogenide semiconductors}
\author{Parijat Sengupta}
\affiliation{Department of Electrical and Computer Engineering, Purdue University, West Lafayette, IN 47907.}
\affiliation{Network for Computational Nanotechnology, Purdue University, West Lafayette, IN 47907.}
\affiliation{Photonics Center, Boston University, Boston, MA 02215.}

\begin{abstract}
We study the low temperature thermal conductivity of single-layer transition metal dichalcogenides. In the low temperature regime where heat is carried primarily through transport of electrons, thermal conductivity is linked to electrical conductivity through the Wiedemann-Franz law. Using a \textit{k.p} Hamiltonian that describes the $ K $ and $ K^{'} $ valley edges, we compute the zero-frequency electric (Drude) conductivity using the Kubo formula to obtain a numerical estimate for the thermal conductivity. The impurity scattering determined transit time of electrons which enters the Drude expression is evaluated within the self-consistent Born approximation. The analytic expressions derived show that low temperature thermal conductivity 1) is determined by the band gap at the valley edges in monolayer TMDCs and 2) in presence of disorder which can give rise to the variable range hopping regime, there is a distinct reduction. Additionally, we compute the Mott thermopower and demonstrate that under a high frequency light beam that sets up a Floquet Hamiltonian, a valley-resolved thermopower can be obtained. A closing summary reviews the implications of results followed by a brief discussion on applicability of the Wiedemann-Franz law and its breakdown in context of the presented calculations. 
\end{abstract}
\maketitle

\vspace{0.35cm}
\section{Introduction}
\vspace{0.35cm}
Transition metal dichalcogenides (TMDCs) which have the representative formula MX$_{2}$ where \textit{M} is a transition metal element from group IV-VI and \textit{X} belongs to the group of elements S, Se, and Te (collectively identified as chalcogens) are layered materials of covalently bonded atoms held together by weak van der Waals forces~\cite{wilson1969transition,wang2012electronics}. The underlying arrangement (see Fig.~\ref{crys}) consists of layers of the transition metal atom surrounded by a chalcogen in a trigonal prismatic arrangement\cite{ramakrishna2010mos2} giving the overall crystal a hexagonal or rhombohedral structure. TMDCs are known to exhibit a wide range of behaviour spanning the whole gamut from metals to semiconductors; however, attention has been directed at the recent progress in exfoliation of the layers in a semiconducting and indirect bulk TMDC which yields a layered two-dimensional (2D) configuration. The 2D monolayer TMDC which is direct band gap with remarkably different microscopic attributes~\cite{chhowalla2013chemistry,huang2013metal} compared to their bulk counterparts are being currently pursued for a diverse set of applications~\cite{chhowalla2016two,radisavljevic2011single} that includes harnessing of their optoelectronic and thermoelectric behaviour. Thin films of TMDCs are considered promising thermal materials~\cite{fan2014mos2,huang2013thermoelectric} with the possibility of a large figure of merit, $ ZT = S^{2}\sigma T/\kappa $, where $ S $ is the Seebeck coefficient while $ \sigma $ and $ \kappa $ denote the electrical and thermal conductivity, respectively. 

Classical models describe the temperature $\left(T\right)$ dependence of the heat capacity by the Debye theory which predicts a $ T^{3} $ relation when $ T \ll \Theta_{D} $, the Debye temperature. Deviations from this law, however, have been found~\cite{gopal2012specific} and attributed to electronic excitations close to the Fermi surface. In this regard, we note that the study of thermoelectric behaviour and attendant transport processes, particularly at low-temperatures offer insight into elementary electronic processes that are usually swamped by the interaction of the lattice with the electron cloud in presence of active phonon modes, primarily through electron-phonon coupling. Further, elucidating the underlying behaviour for reliable information on the low-temperature thermal conductivity, a key measure of thermoelectrics, is crucial in driving the design of efficient devices in this regime, such as hot-electron bolometers, self-integrated Peltier cooling engines, and thermopower-assisted fuel cells. It is useful to recall that typically the total thermal conductivity $\left(\kappa\right)$ has an electronic $\left(\kappa_{e}\right) $ and lattice contribution $\left(\kappa_{ph}\right) $ with the former dominant at low-temperatures while a large number of phonons at elevated temperatures carry the heat current and also impede the electronic transport through multiple scattering mechanisms.~\cite{ziman1960electrons} However, in the low temperature limit, in the absence of substantial phonon distribution, heat carrying electrons are scattered primarily by impurities and defects. In what follows, we will ignore any phonon contribution and the possibility of coupling between the vibrational and electronic modes in our analysis to establish the electronic contribution to the low-temperature thermal conductivity of carriers located at the bottom of the conduction band in the vicinity of the high symmetry valley edges, $ K $ and $ K^{'} $, of monolayer TMDCs. 

We employ the Wiedemann-Franz law (WFL) in deriving $\left(\kappa_{e}\right) $ by connecting to the Drude (zero-frequency intra-band) conductivity $\left( \sigma_{D}\right) $ which is determined by a direct application of the standard Kubo formalism. The eigen states (and corresponding eigen functions) for the Kubo calculation are obtained from a \textit{k.p} description of energy states in a monolayer TMDC around the valley edges of the Brillouin zone. The electron scattering time in the Drude conductivity (and specific heat expression) is acquired from the imaginary part of the retarded self-energy of surface disorder and imperfections. The imaginary part is extracted by setting up the retarded Green function in a self-consistent Born approximation (SCBA) framework. Notice that in the low-temperature regime, phonons are suppressed and do not complement the electron scattering, the retarded self-energy contribution therefore solely involves the contribution of impurities and disorder. We primarily find that close to the conduction band edge the thermal conductivity is enhanced for a higher Fermi level and monolayer TMDCs with a smaller band gap. A notable example of intrinsically shrunken band gap because of a stronger spin-orbit splitting is the monolayer TMDC WSe$_{2}$, a consequence of which is diminished Drude conductivity reflected in its low-temperature thermal counterpart. As a useful ancillary result, a straightforward computation of the specific heat is possible by a simple insertion of the thermal conductivity (and electron transit time) in the kinetic theory of electrons.~\cite{liboff2003kinetic} 

While measurements of $ \kappa_{e} $ serve as a useful probe of electronic behaviour and thermal management, an allied complementary quantity, the thermopower, also allows an examination of related transport characteristics. We use the Mott expression for thermopower which is valid in the low temperature regime, $ T \ll T_{F} $, where $ T_{F} $ is the Fermi temperature. The thermopower, in agreement with experimental observation, displays an inverse relationship to thermal conductivity; while the latter reports a reduction with a higher band gap, a drop is noticed in the former. The calculation of thermopower, in particular, is of significant interest as a higher value translates into better thermoelectric devices. Graphene, for instance, has high thermal conductivity but marked by low thermopower (Seebeck coefficient~\cite{seol2010two}) which suggests their non-viability in design of thermoelectrics; however, Buscema \etal were able to demonstrate a high thermopower for monolayer MoS$_{2}$ and further showed their tunability with an external gate field.~\cite{buscema2013large} In this paper, unlike Ref.~\onlinecite{buscema2013large}, an external gate field is not impressed to alter the Fermi position; rather, we irradiate the sample with a high-energy circularly-polarized beam that rearranges the electronic dispersion and the fundamental band gap. A circularly polarized beam gives rise to Floquet states~\cite{tannor2007introduction} which in the \textit{off-resonant} approximation~\cite{kitagawa2011transport} generates frequency- and valley-dependent band gaps. Such tunable band gaps in a laser-driven setup substantially modulates the thermopower.

The calculations presented here are in the low-temperature regime where the Wiedemann-Franz law holds; however possible sources of violation of the WFL exist and we briefly note instances of those in a closing summary. Additionally, the summary also points out the possibility of other methods such as mechanical strain for improved thermoelectric performance.  

\vspace{0.35cm}
\section{Theory}
\label{theo}
\begin{figure}
\begin{subfigure}
\centering
\includegraphics[width=.4\linewidth]{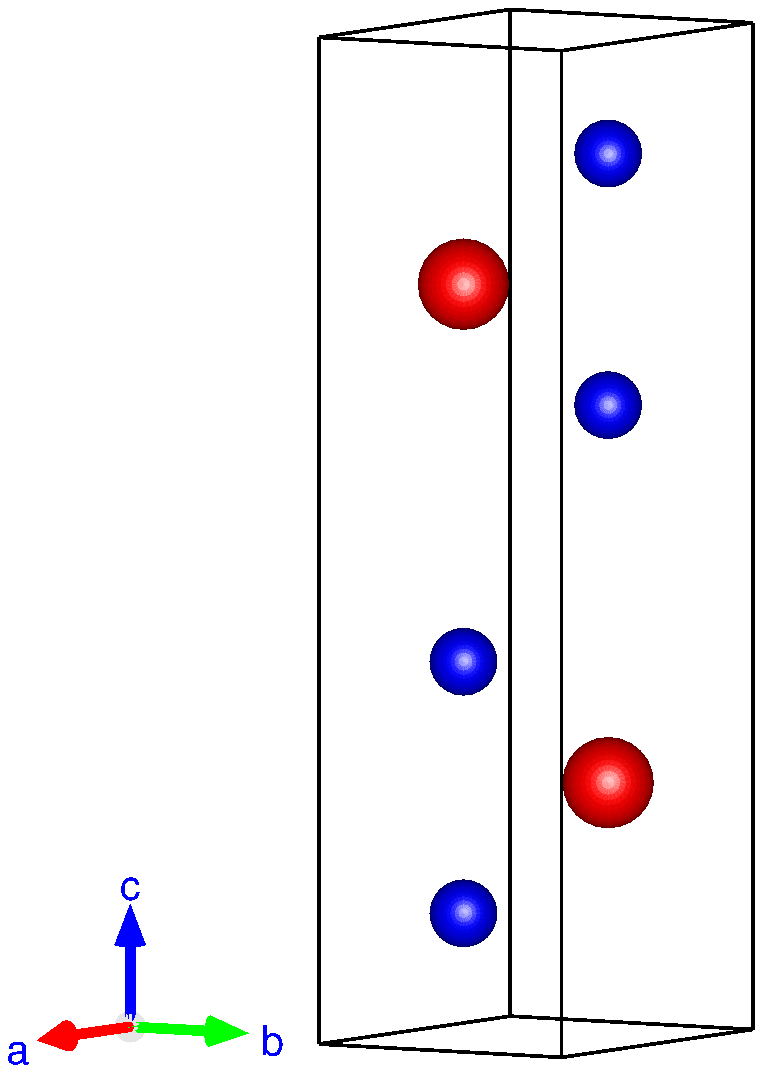}
\end{subfigure}
\begin{subfigure}
\centering
\includegraphics[width=.4\linewidth]{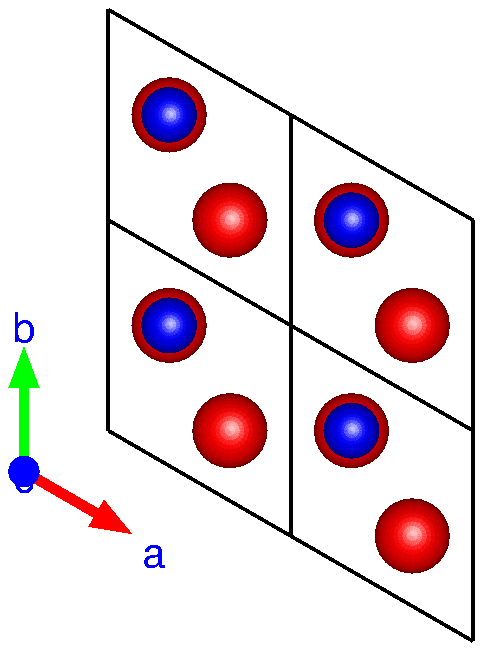}
\end{subfigure}
\caption{The bulk unit cell (left panel) of MoS$_{2}$, a typical semiconducting transition metal dichalcogenide (space group P6$_{3}$/mmc). The two red atoms denote molybdenum (Mo) while four sulfur (S) atoms are shown as blue spheres. Each Mo atom is coordinated to six sulfur atoms in a prismatic fashion. The vertical separation between intra- and inter-layer sulfur atoms is (0.5 - 2z)c and 2zc respectively. For MoS$_{2}$, $ z = 0.12 $ and $ c = 12.58\,\AA $\cite{wang2013mos2}. The right panel is the corresponding top view. The plots for arrangement of atoms were done using the VESTA software\cite{2008vesta}.}
\label{crys}
\end{figure}

\begin{figure}
\centering
\begin{subfigure}
\centering
\includegraphics[width=.7\linewidth]{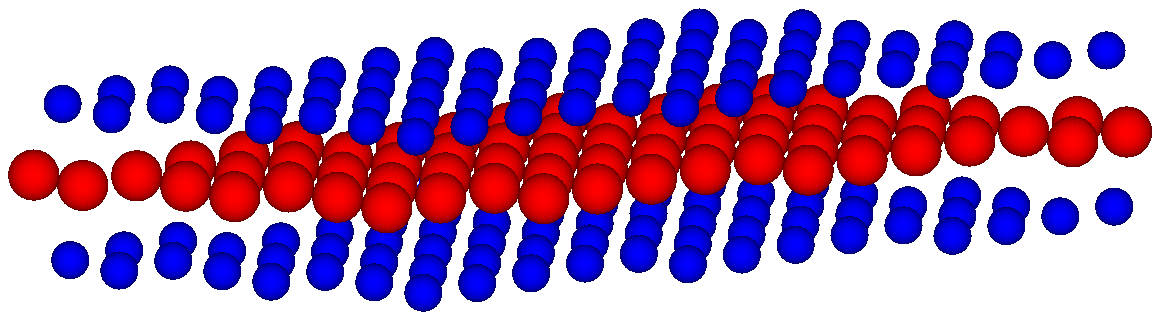} \\
\end{subfigure}
\begin{subfigure}
\centering
\includegraphics[width=.9\linewidth]{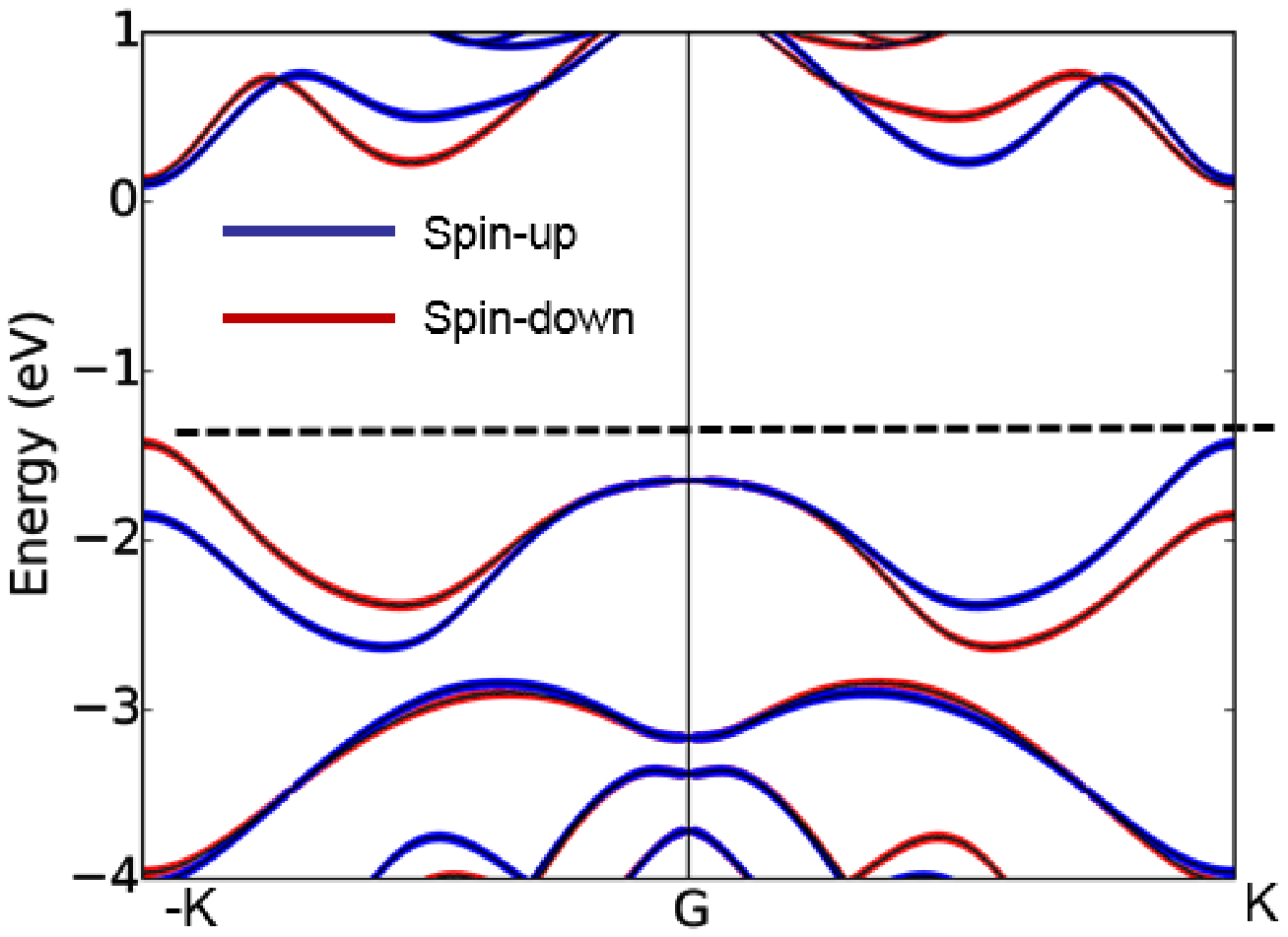}
\end{subfigure}
\caption{The upper panel shows the monolayer of a TMDC (WS$_{2}$ in this case). The metal atom (red) is sandwiched between the sulphur atoms (blue). The tri-layered structure in principle constitutes a single layer for the TMDC. The dispersion of the monolayer was obtained from an ab-initio calculation. The choice of WS$_{2}$ as a representative TMDC is dictated by the fact that it has a significantly large spin-orbit coupling allowing the spin-split bands to be clearly distinguishable. Note the time reversal symmetry mandated flipping of the order of spin bands at the $ K $ and $ K^{'} = - K $ valley edges.}
\label{mldp}
\end{figure}

\vspace{0.35cm}
The basis for all calculations in this paper is the low-energy Hamiltonian shown in Eq.~\ref{mos2ham} for monolayer TMDCs (see upper panel of Fig.~\ref{mldp}). The material constants that appear in the Hamiltonian (Eq.~\ref{mos2ham}) for representative semiconducting TMDCs (see Fig.~\ref{crys}) are listed in Ref.\cite{xiao2012coupled}
\begin{equation}
H_{\tau} = a\,t\left(\tau k_{x}\hat{\sigma}_{x} + k_{y}\hat{\sigma}_{y}\right)\otimes \mathbb{I} + \dfrac{\Delta}{2}\hat{\sigma}_{z}\otimes \mathbb{I} - \dfrac{\lambda\,\tau}{2}\left(\hat{\sigma}_{z}-1\right)\otimes\,\hat{s}_{z}. 
\label{mos2ham}   
\end{equation}
The Hamiltonian in Eq.~\ref{mos2ham} can be split in to a conduction and valence part by expanding the matrices. The 2 $\times $ 2 upper and lower blocks in Eq.~\ref{mos2ham} denote the two sets of spin conduction and valence bands. In this representation, the spin conduction bands are degenerate at the edges while the corresponding valence bands are separated by $ \lambda $, the spin-orbit splitting. For all calculations we use the $ K $ edge and therefore drop the subscript $ \tau $ and set it to unity everywhere. Note that we could have equally worked with the $ K^{'} $ edge $\left(\tau = -1\right)$ which is degenerate with $ K $ and is related to it through time reversal symmetry. To see this explicitly (calculations done with VASP~\cite{kresse1996software}), notice the colour of spin-resolved bands (Fig.~\ref{mldp}) at the $ K $ and $ K^{'} $ edge; the spin-up and spin-down bands interchange order though the fundamental band gaps remains unchanged.  
For later use, we also note the eigen functions and eigen states of the Hamiltonian in Eq.~\ref{mos2ham}. The wave functions at the $ K $ valley edge for the spin-up conduction $\left(+\right)$ and valence $\left(-\right)$ states have the form $\left(\theta = -\tan^{-1}k_{y}/k_{x}\right)$ 
\begin{equation}
\Psi^{up}_{\pm} = \dfrac{1}{\sqrt{2}}\begin{pmatrix}
\eta_{\pm}e^{i\theta} \\
\pm\,\eta_{\mp}
\end{pmatrix},
\label{wfs1}
\end{equation}
where $\eta_{\pm}^{up} = \sqrt{1 \pm \left( \Delta - \lambda\right)/\left(\sqrt{\left(\Delta - \lambda\right)^{2} + \left(2atk\right)^{2}}\right)}$. Note that we can derive an identical set of wave functions for the spin-down components by choosing the lower $ 2 \times 2 $ block. We only need to replace the $ \Delta - \lambda $ in the expression for $ \eta^{up} $ with $ \Delta + \lambda $ to yield the spin-down wave functions. The accompanying eigen functions for the spin-up branch can also be easily written as
\begin{equation}
\varepsilon_{\pm} = \dfrac{1}{2}\biggl[\lambda \pm \sqrt{\left(\Delta - \lambda\right)^{2} + 4a^{2}t^{2}k^{2}}\biggr].
\label{eigval}
\end{equation}
The $ +\left(-\right) $ in the eigen energy expressions correspond to the conduction (valence) band. Note that the finite spin-orbit coupling, $ 2\lambda $, splits the valence bands at $ K $ while the conduction states remain spin degenerate. 

\vspace{0.35cm}
\subsection{Drude Conductivity} 
\vspace{0.35cm}
The main purpose of this letter is the determination of low temperature thermal conductivity of TMDCs via the law of Wiedemann and Franz (WFL). WFL states that if $ \kappa $ is the thermal conductivity ignoring lattice contributions and $ \sigma $ the corresponding electrical conductivity, the ratio $ \kappa/\sigma $ is 
\begin{equation}
\kappa/\sigma = \mathcal{L}T,
\label{wfl}
\end{equation}
where $ \mathcal{L} $ is the Lorentz ratio given as $ \pi^{2}k_{B}^{2}/3e^{2} $ and $ k_{B} $ is the Boltzmann constant. The absolute temperature is $ T $. The electrical conductivity in WFL is the zero-frequency intra-band (Drude) conductity. We evaluate the Drude conductivity using the Kubo expression\cite{bruus2004many} from linear response theory. For a non-interacting sample in 2D space, it is 
\begin{flalign}
\sigma^{\alpha\beta} =  -i\dfrac{\hbar\,e^{2}}{L^{2}}\sum_{n,n^{'}}\dfrac{f\left(\varepsilon_{n}\right)- f\left(\varepsilon_{n^{'}}\right)}{\varepsilon_{n} - \varepsilon_{n^{'}}}\dfrac{\langle\,n\vert\,\hat{v}_{\alpha}\vert\,n^{'}\rangle \langle\,n^{'}\vert\,\hat{v}_{\beta}\vert\,n\rangle}{\varepsilon_{n} - \varepsilon_{n^{'}}+i\,\eta},
\label{kubof} 
\end{flalign}
where $ \vert\,n\rangle $ and $ \vert\,n^{'}\rangle $ are eigen functions (Eq.~\ref{wfs1}) of the Hamiltonian in Eq.~\ref{mos2ham} and $ \eta $ represents a finite broadening (lifetime of the electron on the Fermi surface) of the eigen-states resulting from surface imperfections and impurity scattering. To clarify choice of notation, the superscripts on $ \sigma $ mean that upon application of an electric field along $ \hat{e}_{\beta} $, the electric conductivity tensor gives the current response along $ \hat{e}_{\alpha} $. We have also tacitly assumed that the wave functions retain their pristine form, the presence of impurities and defects notwithstanding. By a direct insertion of the wave functions and the appropriate velocity components in Eq.~\ref{kubof}, we can now determine the longitudinal intra-band conductivity of a monolayer TMDC with sample area $ \mathcal{A} = L^{2} $. The velocity operator along the \textit{x}-axis is $ \hat{v}_{x} = \left(at/\hbar\right)\hat{\sigma}_{x} $. Note that $ \hat{v}_{y} $ is identical since the Hamiltonian is isotropic in the plane. An explicit evaluation of the Drude conductivity begins by setting for the intra-band case, $ \vert\,n\rangle = \vert\,n^{'}\rangle$ in Eq.~\ref{kubof}; the matrix element, $ M = \langle\,n\vert\,\hat{v}_{x}\vert\,n\rangle $, is therefore $ -at\cos\theta\left[2atk/\hbar\left(\sqrt{\Delta_{m}^{2} + \left(2atk\right)^{2}}\right)\right] $. As a shorthand notation, $ \Delta_{m} = \Delta - \lambda $. In obtaining the above expression, we have chosen the conduction band states to evaluate the matrix product; this choice is made by setting the Fermi level to bottom of the conduction band. Inserting the matrix element in Eq.~\ref{kubof}, the Drude conductivity is \begin{equation}
\sigma_{D} = \dfrac{\left(eat\right)^{2}}{4\pi^{2}\hbar\eta}\int_{0}^{2\pi}\cos^{2}\theta d\theta\int kdk\dfrac{\left(2atk\right)^{2}}{\Delta_{m}^{2} + \left(2atk\right)^{2}}\dfrac{\partial f}{\partial \varepsilon}.
\label{dr1}
\end{equation}
In Eq.~\ref{dr1}, we have replaced the summation over momentum states by the integral; additionally the term $f\left(\varepsilon_{n}\right)- f\left(\varepsilon_{n^{'}}\right)/\left(\varepsilon_{n} - \varepsilon_{n^{'}}\right) $ is approximated as $ \partial f/\partial \varepsilon = -\delta\left(\varepsilon_{f} - \varepsilon\right) $ by Taylor expanding the Fermi distribution, $ f\left(\varepsilon_{n^{'}}\right) = f\left(\varepsilon_{n} \right) + \left(\varepsilon_{n^{'}} - \varepsilon_{n}\right)\partial f/\partial \varepsilon $. Note that the relation $ \partial f/\partial \varepsilon = -\delta\left(\varepsilon_{f} - \varepsilon\right) $ holds exactly at $ T = 0 $. Converting the $ k $-space integral to energy space using Eq.~\ref{eigval} and integrating out the angular variable $ \left(\int_{0}^{2\pi}\cos^{2}\theta\,d\theta = \pi\right)$, we rewrite Eq.~\ref{dr1} (normalized to $ e^{2}/h $ ) as 
\begin{equation}
\sigma_{D} = \dfrac{1}{2\eta}\int d\varepsilon\dfrac{\left(2\varepsilon - \lambda\right)^{2}-\Delta_{m}^{2}}{\left(2\varepsilon - \lambda\right)}\delta\left(\varepsilon_{f} - \varepsilon\right) = \dfrac{\Omega}{2\eta},
\label{dr2}
\end{equation}
where $ \Omega = \left[\left(2\varepsilon_{f} - \lambda\right)^{2}-\Delta_{m}^{2}\right]/\left(2\varepsilon_{f} - \lambda\right)$. Writing the broadening parameter, $ \eta = \hbar/\tau_{tr} $ reproduces the expression in form of Drude conductivity. We are now left with the determination $ \eta $ in Eq.~\ref{dr2}; this is obtained from a self-consistent Born approximation (SCBA) outlined in Section.~\ref{scbo}. 

\vspace{0.35cm}
\subsection{Self-consistent Born approximation}
\label{scbo} 
\begin{figure}[!t]
\includegraphics[scale=0.85]{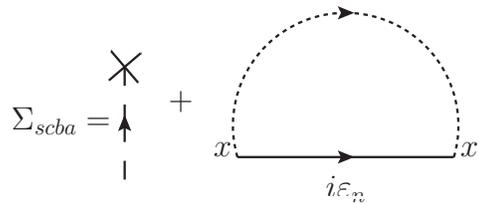} 
\caption{The self energy $ \left(\Sigma_{scba}\right) $ in the Born approximation averaged over impurity distributions. The Matsubara frequency is unchanged since collisions are assumed to be elastic. The dashed line is the average of the two impurity locations marked as $ x $ and $ x^{'} $ while the $ \times $ represents a scattering event.}
\label{feyn1}
\end{figure}

\vspace{0.35cm}
The broadening is considered as arising out of disorder on the surface and is modeled as an effective retarded self-energy within SCBA.~\cite{bruus2004many} The pair of SCBA equations being:
\begin{equation}
\begin{aligned}
\label{scba1}
G_{ks}\left(\epsilon\right) = \dfrac{1}{\epsilon - \epsilon_{ks} - \Sigma\left(\epsilon\right)};
\Sigma\left(\epsilon\right) = n_{i}v_{i}^{2}\int\,\dfrac{d^{2}k}{4\pi^{2}}G_{ks}\left(\epsilon\right),
\end{aligned}
\end{equation}
where $ n_{i} $ and $ v_{i} $ denote the density and strength of impurities, respectively and $ G_{ks}\left(\epsilon\right) $ is the retarded Green's function diagonal with respect to the band index \textit{s} ($ \langle\,s\vert\,G_{k}\left(\epsilon\right)\vert\,s\rangle = \delta_{ss^{'}}G_{ks}\left(\epsilon\right) $). The self-energy $ \Sigma $ which is also diagonal with respect to the band index \textit{s} and independent of \textbf{\textit{k}} in SCBA is averaged over impurity distributions (see Fig.~\ref{feyn1}). The unperturbed retarded Green's function for the  $ 2 \times 2 $ upper block of the Hamiltonian in Eq.~\ref{mos2ham} is $ G_{0, R} = \left(E - H^{2 \times 2} + i\delta\right)^{-1} $. Inserting $ G_{0,R} $ in the self-energy expression (Eq.~\ref{scba1}) and recasting to the form $ \dfrac{1}{x \pm i0^{+}} $ to separate the real and imaginary parts using the standard expression $ \dfrac{1}{x \pm i0^{+}} = \mathbb{P}\dfrac{1}{x} \mp i\pi\delta\left(x\right) $, we approximately arrive at: 
\begin{flalign}
Im\,\Sigma &= n_{i}v_{i}^{2}\int \dfrac{d^{2}k}{4\pi^{2}}\biggl[\delta\left(E + \Delta/2 - \lambda\right) + \delta\left(E - \Delta/2\right)\biggr], \notag \\
&\approx n_{i}v_{i}^{2}\dfrac{1}{2a^{2}t^{2}}\left(\Delta/2\right).
\label{imself2}
\end{flalign}
The imaginary retarded self-energy term is linked to scattering time, $\tau_{tr} $, by the relation $ \hbar/\tau_{tr} = 2Im\Sigma $. The real part simply of the self-energy renormalizes the Fermi energy and is absorbed in the chemical potential. We have neglected the $ a^{2}t^{2}k^{2} $ terms in Eq.~\ref{imself2} since close to the valley edge $ k $ is a small number and the product $ atk $ can be ignored. Notice that the energy arguments of the two $ \delta\left(\cdot\right) $ functions in Eq.~\ref{imself2}, $ \lambda - \Delta/2 $ and $ \Delta/2 $, happen to be aligned to the top and bottom of the valence and conduction band, respectively. Since we carry out calculations around the conduction band minimum, the argument $ \lambda - \Delta/2 $ is discarded. 

\vspace{0.35cm}
\section{Thermal conductivity and thermopower} 
\vspace{0.35cm}
From the general expression for the Drude conductivity, the low temperature thermal conductivity $\left(\kappa_{e}\right)$ ignoring phonon contribution can be established by a simple application of the Wiedemann-Franz law (WFL) as briefly noted (Eq.~\ref{wfl}) in the preceding section. A correct application of WFL is incumbent on weak elastic scattering of electrons and negligible electron-electron correlation, \ie the electrons move independent of one another. Assuming that the ensemble of electrons for a monolayer TMDC located in the vicinity of the conduction band minimum fulfill the criteria set forth by WFL, we simply substitute the Drude conductivity expression from Eq.~\ref{dr2} in Eq.~\ref{wfl} to obtain $ \kappa_{e} $. The expression takes the form
\begin{equation}
\kappa_{e} = \dfrac{\pi^{2}k_{B}^{2}T}{3e^{2}}\dfrac{\Omega}{2\eta}.
\label{kappawfl}
\end{equation}
The thermal conductivity from Eq.~\ref{kappawfl} evidently depends on the broadening parameter $\left(\eta\right)$ since it directly controls the electric conductivity. We have ignored any correction to the conductivity, however, arising from any weak localization present on the surface due to the assumed impurity concentration. The WFL has been verified for numerous cases and has been proven correct and is generally regarded as a defining proof of the Fermi liquid theory of electrons. Violations to WFL exist (we discuss that in the summary section) but for our purpose where we apply it to a non-interacting body of electrons in monolayer TMDCS, it should suffice. Since thermal conductivity obtained with WFL is directly proportional to electric conductivity for a given temperature, we may easily infer that $ \kappa_{e} $ in monolayer TMDCs will exhibit the same trend as $ \sigma $, the Drude conductivity. 

Here we make note of a useful result on specific heat, a quantity that can be directly measured and is easily determined from the preceding thermal conductivity calculation. The kinetic theory of electron transport relates the thermal conductivity and specific heat as:
\begin{equation}
\kappa_{e} = \dfrac{1}{3}C_{e}v_{f}\Lambda_{e}.
\label{wflc}
\end{equation}
In Eq.~\ref{wflc}, the Fermi velocity is $ v_{f} $, the mean free path is $ \Lambda_{e} $, and $ C_{e} $ is the specific heat of electrons. Note that an analogous relation for the phonon contribution to overall thermal conductivity at elevated temperatures (when the phonon population is significant) exists but we ignore it here. The mean free path $ \Lambda_{e} = v_{f}\tau $, where $ \tau $ is the relevant scattering time.

\vspace{0.35cm}
\subsection{Mott's expression for TMDC and laser-driven thermopower} 
\vspace{0.35cm}
Analogous to thermal conductivity calculations, we can also determine the thermopower $ \left(\mathcal{Q}\right) $ of a monolayer TMDC via the Mott formula, which is
\begin{equation}
\mathcal{Q} = -\dfrac{\pi^{2}}{3e}\dfrac{k_{B}^{2}T}{\sigma}\dfrac{\partial \sigma}{\partial \varepsilon}.
\label{mott}
\end{equation}
Inserting Eq.~\ref{dr2} in Eq.~\ref{mott} and the expression for the derivative, the thermopower expression simplifies to
\begin{equation}
\begin{aligned}
\mathcal{Q} &= -\dfrac{\pi^{2}}{3e}\dfrac{2\eta\,k_{B}^{2}T\left(2\varepsilon_{f} - \lambda\right)}{\left(2\varepsilon_{f} - \lambda\right)^{2}-\Delta_{m}^{2}}\dfrac{\partial \sigma}{\partial \varepsilon}, \\
& = -\dfrac{2\pi^{2}}{3e}k_{B}^{2}T\dfrac{1+t}{1-t}\dfrac{\sqrt{t}}{\Delta_{m}}.
\label{motttm}
\end{aligned}
\end{equation}
In Eq.~\ref{motttm}, $ \sqrt{t} = \Delta_{m}/\left(\left(2\varepsilon_{f} - \lambda\right)\right) $. It is worthwhile to mention that the Mott thermopower expression holds good insofar as the approximation of representing the Fermi distribution as a step function. For cases, where considerable smearing of the bands is present, a large deviation between the result contained in Eq.~\ref{motttm} and experimental data must be expected.

It is apparent from Eq.~\ref{motttm} that the overall band gap $ \left(\Delta_{m}\right) $ influences the Mott thermopower. In connection to the applicability of this result to the field of thermoelectrics at the nanoscale, it would be prudent to consider an approach that allows a measure of external control by virtue of alteration to the band energy description. In light of this, we examine the possibility of laser-driven periodic perturbation that engineers the energy dispersion of a monolayer TMDC. A periodic perturbation in quantum mechanics is dealt by invoking the Floquet theory that allows the construction of an effective time independent Hamiltonian. The theory is summarized in several published works.~\cite{kitagawa2011transport,cayssol2013floquet,lopez2015photoinduced} We simply quote the result here that shows the change to the band gaps at the $ K $ and $ K^{'} $ edges when placed under a high-frequency light source, commonly known in literature as the \textit{off-resonant} condition. 

\begin{figure}
\includegraphics[scale=0.65]{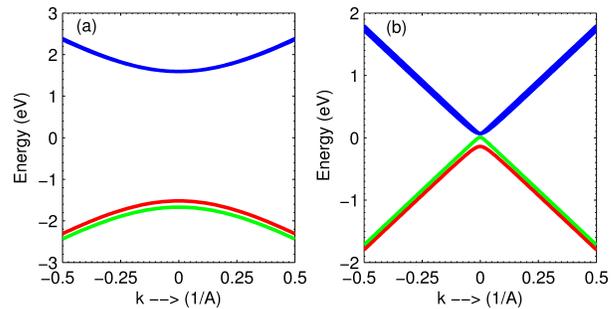}
\caption{The dispersion of monolayer MoS$_{2}$ under \textit{off-resonant} light condition. The sub-figure on the left (right) plots the band dispersion around the $ K (K') $ point. The energy of the light beam was assumed to be $ eAv_{f} = 2.9\, eV $.  This result is in qualitative agreement with Ref.~\onlinecite{tahir2014photoinduced,sengupta2016photo}.}
\label{disp_altered}
\end{figure}

The influence of the periodic \textit{off-resonant} light on the TMDC monolayer is to the lowest order approximated by an effective Hamiltonian averaged over a complete cycle through the evolution operator $ U = \mathcal{T}exp\left(-i\int_{0}^{T}H\left(t\right)dt\right)$.~\cite{kitagawa2011transport} Here $ \mathcal{T} $ is the time-ordering operator and $ T = 2\pi/\omega $. This approximate Hamiltonian, which in principle describes the behaviour of a system with time scales much longer than $ T $, rearranges the electron occupation number without modifying the bands. In the \textit{off-resonant} state, the approximate Floquet Hamiltonian following Ref.~\onlinecite{kitagawa2011transport} is
\begin{equation}
H_{\mathcal{F}} = H_{\tau} + \dfrac{1}{\hbar\,\omega}\left[H_{-1}, H_{1}\right],
\label{flham}
\end{equation}
and $ H_{m} = \dfrac{1}{T}\int_{0}^{T}H\left(t\right)exp\left(-im\omega t\right)dt $. Note that $ H\left(t\right) $ is the time-dependent part obtained using the standard Peierl's substitution $ \hbar\,k\rightarrow \hbar\,k - e\textbf{A}\left(t\right)$ in the TMDC monolayer Hamiltonian (Eq.~\ref{mos2ham}); this substitution gives $
H\left(t\right) =  \dfrac{at}{\hbar}\,A\left(\sigma_{x}cos\,\omega t + \sigma_{y}sin\,\omega t\right) $, where the \textit{off-resonant} light is right-circularly polarized and represented through the vector potential $ \textbf{A}\left(t\right) =  A\left(cos\,\omega t\,\hat{e}_{x}, sin\,\omega t\,\hat{e}_{y}\right) $. The amplitude and frequency are denoted by $ A $ and $ \omega $, respectively. The desired Floquet Hamiltonian, $ H_{\mathcal{F}} $, by a direct evaluation of the respective Fourier components and using $ \left[\sigma_{x}, \sigma_{y}\right] = 2i\sigma_{z} $ therefore reads similar to Eq.~\ref{mos2ham} but with a different band gap. The change in band gap by evaluating the commutator in Eq.~\ref{flham} and inserting in Eq.~\ref{mos2ham} is expressed as $ \Delta_{m}\sigma_{z}\otimes\mathbb{I}\rightarrow \left(\Delta_{m} + \tau\,\Delta_{F}/2\right]\sigma_{z}\otimes\mathbb{I} $, where the Floquet-induced band gap modification gap is
\begin{equation}
\Delta_{F} = 2e^{2}A^{2}a^{2}t^{2}/\hbar^{3}\omega.
\label{flbg}
\end{equation}
In Eq.~\ref{flbg}, $ A = E_{0}/\omega $ where $ E_{0} $ is the amplitude of the electric field. A more convenient representation utilizing the relation $ at = \hbar\,v_{f} $ allows us to write this as $ 2\left(eAv_{f}\right)^{2}/\hbar\omega $. This light-induced band gap under \textit{off-resonant} conditions is alterable through the intensity and frequency parameters by expressing the intensity of incident light as $ I = \left(eA\omega\right)^{2}/\left(8\pi\alpha\right)$, $ \alpha = 1/137 $ being the fine structure constant.~\cite{cayssol2013floquet} The Floquet modulated band gap is therefore $ 16\pi\alpha\,Iv_{f}^{2}/\omega^{3} $. The dispersion diagram when right-circularly polarized light (under \textit{off-resonant} conditions) shines on a monolayer of MoS$_{2}$ with altered band gaps is shown in Fig.~\ref{disp_altered}. Notice that the band gap at $ K $ is increased to $ 3.11\, eV $ from the pristine $ 1.66\,eV $ while its time-reversed counterpart at $ K^{'} $ sees a reduction to $ 0.074\,eV $ for right-circularly polarized light. The enhancement and reduction at the valley edges is reversed for a left-circularly polarized beam. The new band gap $ \left(\Delta_{m} + \Delta_{F}/2\right) $ can be substituted in Eq.~\ref{motttm} to obtain a driving frequency-reliant thermopower.

\vspace{0.35cm}
\subsection{Thermal power in the variable range hopping regime} 
\vspace{0.35cm}
In an earlier section, the use of SCBA in presence of impurity disorder supplied us with a finite broadening of the density of states; however, material constants were left unchanged, a tacit set of assumptions that isn't necessarily true. Disorder-induced localization, in addition to serving as an agent for tangible changes to electron transport also reduces the electrostatic screening to enhance the long-range Coulombic interaction and rearranges the distribution of energy states, a clear expression of which is mirrored in a changed set of material parameters. While the intrinsic spin-orbit coupling, a key material parameter in monolayer semiconducting TMDCs, is normally invariant and unlikely to be influenced through external perturbations, numerical calculations do show that $ `t' $, the hopping parameter (see Eq.~\ref{mos2ham}) can indeed be altered. As a matter of fact, strain, embedded impurities, positional disorder etc., all of which have been shown to be present on the surface of a monolayer TMDC can contribute to the probability of altered hopping.~\cite{qiu2013hopping} A quantitative assessment of their influence can be gauged from the empirical relationship for the probability of electron hopping in a disordered 2D system.~\cite{tessler2009charge} The model uses the electron wave function localization for a specific disorder strength and the hopping radius to predict the following expression
\begin{equation}
P \sim \exp\left(-\dfrac{2R}{\xi} - \dfrac{1}{\pi R^{2}D\left(E\right)kT}\right),
\label{prob} 
\end{equation} 
where $ \xi $ is the localization length, $ R $ is the hopping radius, and $ D\left(E\right) $ is the density of states. The genesis of Eq.~\ref{prob} lies in Mott's variable range hopping (VRH) model; this model advanced by Mott contends that at low temperatures an electron does not always hop to the nearest neighbour but to a state with the lowest activation energy and the shortest hopping distance. For an optimum hopping distance $ `r' $, the maximum hopping probability is expressed by Eq.~\ref{prob}. Since the electric conductivity is linked to the strength of the hopping parameter $ `t' $ which undergoes an adjustment in the Mott model, the thermal conductivity in the WFL regime  must therefore manifestly exhibit an identical behaviour. It can be shown~\cite{mott1993conduction,park2015hopping} that a functional dependence of the electrical conductivity within the Mott-VRH framework can be expressed as
\begin{equation}
\sigma_{\alpha\beta} = \sigma_{\alpha\beta}^{0}\exp\left(-\dfrac{\Lambda}{T}\right)^{\nu},
\label{condt}
\end{equation}   
where $ \nu = 1/3 $ for 2D systems, $ \sigma_{\alpha\beta}^{0} $ is the conductivity at $ T = 0 $ and $ \Lambda $ is an experimentally determined constant, dependent on the radius of hopping/localization length and the density of states close to the Fermi level. Numerical results for thermal conductivity and thermopower centred around the expressions derived here are presented in Section.~\ref{s3}.

As a useful addendum to the thermal conductivity calculations, it is also possible - bearing in mind the preceding discussion on the variability of electric conductivity in presence of disorder and other imperfections - to express the thermal power density of a system at low temperatures as a function of material constants. To carry out this task, we write down the standard heat equation
\begin{subequations}
\begin{equation}
Q_{h} = \kappa_{e}A\dfrac{dT}{dx},
\label{hq1}
\end{equation}
where $ Q_{h} $ is the heat flowing through the system, $ A $ is the area of cross-section, and $ dT/dx $ is the temperature gradient. Substituting for $ \kappa_{e} $ using the WFL (Eq.~\ref{wfl}) in Eq.~\ref{hq1} and the modified conductivity (Eq.~\ref{condt}) gives
\begin{equation}
\int_{0}^{x}\mathcal{P}dx = L\sigma_{\alpha\beta}^{0}\int_{T_{c}}^{T}\exp\left(-\dfrac{\Lambda}{T}\right)^{1/3}T dT.
\label{hq2}
\end{equation} 
The power density is denoted as $ \mathcal{P}\left(x\right) = Q_{h}/A $ and $ T_{c} $ is the constant temperature maintained at one end of the channel. For no spatial dependence of the power density $ \mathcal{P} $, Eq.~\ref{hq2} is
\begin{equation}
\mathcal{P} = \dfrac{L\sigma_{\alpha\beta}^{0}}{l}\int_{T_{c}}^{T}\exp\left(-\dfrac{\Lambda}{T}\right)^{1/3}T dT,
\label{hq3}
\end{equation}
where $ l $ is the length of the one-dimensional channel of heat flow. The indefinite integral on the R.H.S can be recast as $ \mathcal{P} = -3\Lambda^{2}\dfrac{L\sigma_{\alpha\beta}^{0}}{l}\int z^{-7}\exp\left(z\right)dz $, where $ z = -\Lambda/T $. This integral can be either numerically evaluated or using the relation $ \int \dfrac{\exp\left(z\right)}{z^{n}}dz = \dfrac{1}{n-1}\biggl(-\dfrac{\exp\left(z\right)}{z^{n-1}} + \int \dfrac{\exp\left(z\right)}{z^{n-1}}dz\biggr) $ analytically determined through successive integration by parts to furnish the power density.    
\end{subequations}
 
\vspace{0.35cm} 
\section{Numerical results}
\label{s3} 
\begin{figure}
\includegraphics[scale=0.75]{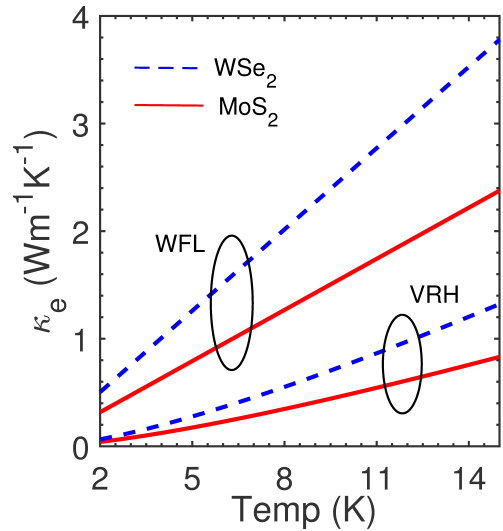}
\caption{The low temperature thermal conductivity for two cases is shown. The curves marked as `WFL' are obtained by a straightforward application of the Wiedemann-Franz law; the other group denoted by `VRH' pertains to the state when variable range hopping is active and modifies the result of WFL as explained in the text (see Eq.~\ref{condt}). The two semiconducting TMDCs are MoS$_{2}$ and WSe$_{2}$ (dashed line).}
\label{therm}
\end{figure}

\vspace{0.35cm}
We have now gathered all the information for a quantitative determination of the low-temperature thermal conductivity $\left(\kappa\right)$. As a first step, we use Eq.~\ref{imself2} to obtain the energy broadening of the states; setting the impurity concentration to $ 2.5 \times 10^{10}\,cm^{-2} $ and attendant impurity potential~\cite{adam2009theory} as $ 0.1\, keV\,\AA^{2} $, the imaginary contribution of the self-energy is approximately equal to $ 4.6\, meV $ and $ 8.0\, meV $ for WSe$_{2}$ and MoS$_{2}$, respectively. A plot of $ \kappa_{e} $ for these two TMDCs as a function temperature for electrons in the vicinity of the bottom of conduction band (note that the conduction band minimum in each case is $ \Delta/2 $, where $ \Delta $ is the fundamental gap at the valley edges) is displayed in Fig.~\ref{therm}. We wish to point out that of the two semiconducting monolayer TMDCs chosen, MoS$_{2}$ and WSe$_{2}$, the latter has greater thermal conductivity at low-temperatures. This is in agreement with their intrinsic Drude conductivities; observe from Eq.~\ref{dr2} that a lower band gap translates into higher Drude conductivity which is the case with WSe$_{2}$. To understand this better, the material parameters in Table.~\ref{table1} reveals a nearly three-fold larger spin-orbit splitting $\left(\lambda\right)$ in WSe$_{2}$ in comparison to MoS$_{2}$ while the other parameters are nearly identical. This large spin splitting (on account of the heavier metal, tungsten) effectively contracts the band gap $\left(\Delta - \lambda\right) $ from which the pattern displayed by the Drude and thermal conductivity (in Fig.~\ref{therm}) follows. There are other physical situations, for instance, the strength of inter-band (valence to conduction state jumps) tranistion rates where the lower band gap of WSe$_{2}$ would again appear as a determining factor; we do not consider such cases here, for a clear example of this, see Ref.~\onlinecite{sengupta2016intensity}. 

When variable range hopping dominates with electrons close to the Fermi level hopping from one localized site to another, the adjusted conductivity (Eq.~\ref{condt}) is pared, an illustration of which is the degrading of attendant thermal conductivity in Fig.~\ref{therm}. For a numerical calculation, the constant $ \Lambda $ was set to $ 17.4\, K $ obtained from a fitting analysis presented in Ref.~\onlinecite{wu2014large} for temperatures under $ 20\, K $. A lower thermal conductivity in absence of pristine crystalline order such as in a nanocrystal may be of value in applications that target thermopower generation. A more detailed note on this point appears in Section.~\ref{summ}.

The quantitative determination of the thermal conductivity, by virtue of Eq.~\ref{wflc} also permits an estimation of the specific heat. For a numerical answer, we assign values to the following quantities: the thermal conductivity for a 1.0 cm$^{2}$ sample of MoS$_{2}$ monolayer (which is $ 6.0\, \AA $ thick\cite{splendiani2010emerging}) is set to $ \kappa_{e} = 1\, W/m\,K $, the transit time using the imaginary part of the self-energy computed (the imaginary retarded self-energy term is linked to scattering time, $\tau_{sc} $, by the relation $ \hbar/\tau_{sc} = 2Im\Sigma $) above is roughly $ 2.0 ps $ and the Fermi velocity $ \left(at/\hbar\right) $ is given a value of $ 5.33 \times 10^{5}\, m/s $. Inserting all of them in Eq.~\ref{wflc}, the specific heat for the MoS$_{2} $ slab is $ 3.2 \times 10^{5}\, eV/K $. The temperature for this calculation was set to $ 10 \, K $. Notice that is the specific heat at low-temperatures ($ T \ll \Theta_{D} $) where electrons primarily carry the heat and lattice contribution via dominant phonon modes is negligible. 
\begin{table}[!b]
\caption{Band structure parameters for monolayer TMDCs~\cite{xiao2012coupled}.}
\centering
\label{table1}
\begin{tabular}{lcccc}
\noalign{\smallskip} \hline \hline \noalign{\smallskip}
Parameters & MoS$_{2}$  & WSe$_{2}$ \\\hline
a(\AA) & 3.193 & 3.310  \\
$ \Delta\,(eV) $ & 1.66 & 1.60  \\
$ t\,(eV) $ & 1.10 & 1.19 \\
$ 2\lambda\,(eV) $ & 0.15 & 0.46 \\
\noalign{\smallskip} \hline \noalign{\smallskip}
\end{tabular}
\end{table}

We next turn our attention to low-temperature thermopower result derived (Eq.~\ref{motttm}) using the Mott formula. First of all, note that the thermopower (or the Seebeck coefficient) exhibits a dependence on the intrinsic band gap $ \left(\Delta - \lambda\right) $ and is independent of the broadening parameter $ \left(\eta\right)$. Indeed, a comparison of $ \mathcal{Q} $ in MoS$_{2}$ and WSe$_{2}$ shows it to be higher for a range of energies in the vicinity of the top of the valence band (Fig.~\ref{therm}). The calculation was done at $ T = 10\,K $. While we show the variation of $ \mathcal{Q} $ for two pristine semiconducting TMDCs here, for an enhanced low-temperature thermopower, methods that could possibly adjust (and lower) the band gap therefore are of interest. In this regard, it will be useful to mention that it is now also possible to synthetically fabricate (apart from exfoliation) single layer alloys of TMDCs. J. Mann \etal report in Ref.\cite{mann20142} the fabrication of single layer Mo$_{1-x}$W$_{x}$S$_{2}$ and MoSe$_{2(1-x)}$S$_{2x}$ allowing for a continuous tuning of the band gap and optical properties by varying the alloy composition. The direct band gap of the alloy, MoSe$_{2(1-x)}$S$_{2x}$, for example, can lie between  1.66 $\mathrm{eV} $ (MoS$_{2}$) and 1.47 $\mathrm{eV} $ (MoSe$_{2}$), assuming the rule of virtual crystal approximation is reasonably valid. 

In passing we note that the expression for thermopower $ \left(\mathcal{Q}\right) $ in Eq.~\ref{motttm} shows a functional independence the broadening parameter. It is reasonable to expect, however, that surface impurities and dopants will influence the thermopower generated; this apparent non-dependence can be explained by noting that $ \eta $ is an energy-independent quantity that we obtained from a self-consistent Born approximation by assigning an impurity concentration and potential in the dilute limit. In a real experimental setup, the broadening parameter $ \eta = \hbar/\tau $, ($ \tau $ is the transit time) is not a fixed quantity and must change as a function of carrier energy. An alternative approach to incorporate the energy dependence would be to use an expression for conductivity in the diffusive limit; in equation form, it should read as
\begin{equation}
\sigma = \Phi D\left(\varepsilon\right)\tau\left(\varepsilon\right)/2.
\label{condiff}   
\end{equation}
This conductivity expression ($\Phi $ is material dependent and D($\varepsilon$) is the density of states) can now be inserted in Eq.~\ref{motttm} for an evaluation of the thermopower. In general, as was shown by Hwang \etal in Ref.~\onlinecite{hwang2009theory}, the energy dependence can be of the form $ \tau \propto \varepsilon^{m} $ with varying values of $ m $ corresponding to different scattering mechanisms. We have only considered an energy-independent impurity scattering here.

\begin{figure} 
\includegraphics[scale=0.75]{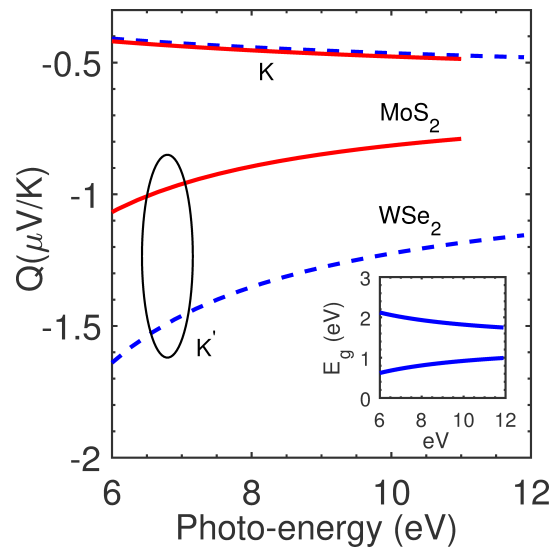}
\caption{The numerically computed valley-resolved low temperature thermopower $ \left(\mathcal{Q}\right)$ of monolayer TMDCs MoS$_{2}$ and WSe$_{2}$ under a high frequency right-circularly polarized light beam is shown. The temperature was set to $ T = 10\, K $. Under \textit{off-resonant} conditions, the enlargement of the band gap at $ K $ provides a higher $ \mathcal{Q} $ compared to its time reversed counterpart at $ K^{'} $. In general, the intrinsically lower band gap for WSe$_{2}$ is expected to provide a smaller $\mathcal{Q} $. The inset shows the progression of the band gaps at the $ K $ and $ K^{'} $ valley edges with incident right-circularly polarized light.}
\label{floq}
\end{figure}

In the last section, we quantitatively show the influence of the \textit{off-resonant} circularly polarized light that introduces a photo-induced energy band gap through the Floquet dressed states. In the brief discussion presented in Section.~\ref{theo}, the band gaps at the time reversed $ K $ and $ K^{'} $ valleys were enlarged and shrunken by shining right-circularly polarized light (see Fig.~\ref{disp_altered}) which in principle could regulate the thermopower, a gap-dependent quantity. Plugging in the altered band gaps in Eq.~\ref{motttm}, we plot (Fig.~\ref{floq}) the photo-controlled thermopower for a range of frequencies. The thermopower follows the well-defined trend and exhibits an upward tick when the band gap is increased. For our case, under a right-circularly-polarized light beam, the band gap at $ K $ is higher than its intrinsic value and therefore furnishes a higher thermopower while the reduced band gaps at $ K^{'} $ for both WSe$_{2}$ and MoS$_{2}$ display a correspondingly lower value. At the $ K^{'} $ edge, the band gap reduction is smaller for MoS$_{2}$, which indicates a higher thermopower over WSe$_{2}$. Notice that the frequencies must satisfy the condition, $ \hbar\omega \gg H $, that is energy contained in the incident beam is far greater than the energy scales of the static problem (typified in the Hamiltonian, $ H $). 

In passing we note that as a matter of fact, in graphene, the earliest 2D material, such \textit{off-resonant} conditions have been fulfilled by using photon energies that lie in the soft X-ray regime. By simulating identical conditions in monolayer TMDC which can be likened to spinful gapped graphene, the thermopower plot (Fig.~\ref{floq}) shows a clear enhancement in case of the $ K^{'} $ valley which has a reduced band gap in contrast to the $ K $ valley edge; the change tailing off as the frequency of the incident light increases. While we have demonstrated a valley-resolved thermopower with right-circularly polarized light, note that results do not qualitatively change under a left-circularly polarized light; as opposed to an enhancement at $ K^{'} $, the $ K $ valley edge now exhibits the same trend. This is simply a consequence of the time reversal symmetry that exists in the system. In any case, regardless of the chirality of the irradiating beam, the valley-resolved thermopower is maintained.

\vspace{0.3cm}
\section{Summary}
\label{summ}
\vspace{0.3cm}
In this work we have carried out an evaluation of the low-temperature thermal conductivity of monolayer semiconducting transition metal dichalcogenides (TMDCs). We calculated the Drude conductivity and related it to the low-temperature thermal conductivity using the Wiedemann-Franz law. Specifically, we established the dependence of thermal conductivity and thermopower (Seebeck coefficient) on the dispersion of the monolayer TMDC. TMDCs with higher band gaps have a larger Seebeck coefficient (and a lower thermal conductivity) which is further tunable under a high-energy circularly-polarized light beam. However, an important remark about the results derived is in order; most importantly, we have tacitly assumed a free-standing monolayer of TMDC while experimental setups may utilize a substrate. The presence of a substrate can modify the results, at least quantitatively, an instance of which can be found in theoretical results reported in Ref.~\onlinecite{babaei2014large}. Substrate-grown monolayer MoS$_{2} $ sheets revealed a poor thermoelectric power factor over their freely-suspended counterparts. 

A chief purpose of this work was to present a description of conditions that are promising to easily adjust the thermal conductivity and thermopower for a wide spectrum of applications.~\cite{wang2012electronics,jariwala2014emerging} In this regard, we note that applications that desire a faster transport of heat to lower ambient temperatures, such as in nano-sized devices that suffer from self-heating, the mechanism of Peltier cooling with a higher thermal conductivity is a necessary condition; on the contrary thermoelectric power generation needs a larger thermopower/Seebeck coefficient. Notice that the two quantities exhibit opposite trends with respect to the intrinsic band gap. Lastly, it is useful to remark that we suggested a laser-driven tuning of the band gap and the thermoelectric behaviour; in addition, optimally straining the monolayer TMDC can also yield the sought characteristics. A promising thermoelectric figure of merit $ \left( ZT \right) $ in case of strained ZrS$_{2} $ monolayer has already been achieved.~\cite{lv2016strain} 

TMDC films also carry defects, vacancies, clusters, and dislocations from the growth process which affect the electronic and chemical behaviour; testimony to which lies in the large body of work/data available from optoelectronic characterization of TMDCs thin films. These measurements clearly show the presence of defect-induced traps that give rise to additional photoemission peaks and distinct photoluminescence intensity.\cite{tao2014strain,mouri2013tunable} These imperfections, rather than being severely detrimental to their device prospects can be turned in to efficient `knobs' by leveraging their influence on the overall thermal attributes of TMDCs; the thermal conductivity, in fact, has been shown to be modulated through defects in silicene\cite{li2012vacancy,liu2014thermal}. We utilized Mott's variable-range-hopping model to describe the change in Drude conductivity and how a modulation of its thermal equivalent could be accomplished. It is pertinent to state that the analysis presented here involves a Hamiltonian (Eq.~\ref{mos2ham}) that describes massive Dirac fermions around the valley edges; in principle this study could also be extended~\cite{rostami2014intrinsic} to gapped topological insulators. Such topological insulators host massive Dirac fermions and the gap opening of the surfaces states could be a result of inter-surface hybridization in thin films or the presence of an out-of-plane magnetic field. 
 
Before closing, we wish to remark about the validity of using the Wiedemann-Franz law (WFL) to calculate the thermal conductivity. WFL tacitly assumes that the ensemble of electrons do not undergo inelastic electron-phonon scattering and electron-electron interaction is negligible. However, violations to WFL appear for strongly interacting systems such as heavy fermion metals,\cite{tanatar2007anisotropic} Luttinger liquids, and ferromagnets and is normally considered as the hallmark of non-Fermi liquid behaviour. A recently reported work\cite{crossno2016observation} also predicts the violation of WFL in two-dimensional graphene in vicinity of the charge neutral point that hosts a quasi-relativistic electron-hole plasma known as the Dirac fluid. 


\end{document}